\documentclass{optica-article}

\journal{opticajournal} 
\articletype{Research Article}

\usepackage{lineno}

\begin{document}

\title{Moment-based Space-variant Shack–Hartmann Wavefront Reconstruction}

\author{Fan Feng\authormark{1,2,*,$ \dagger $}, Chen Liang,\authormark{3,$ \dagger $} and Dongdong Chen\authormark{4}, Ke Du\authormark{5}, Runjia Yang\authormark{3}, Chang Lu\authormark{6}, Shumin Chen\authormark{4}, Wenting He\authormark{7}, Pingyong Xu\authormark{7,8}, Liangyi Chen\authormark{5,9,10}, Louis Tao\authormark{
1,11}, Heng Mao\authormark{3,12,13}}

\address{\authormark{1}Center for Bioinformatics, National Laboratory of Protein Engineering and Plant Genetic Engineering, School of Life Sciences, Peking University, Beijing 100871, China\\
\authormark{2}GBA Institute of Collaborative Innovation, Guangzhou 510300, China\\
\authormark{3}LMAM, School of Mathematical Sciences, Peking University, Beijing 100871, China\\
\authormark{4}School of Software and Microelectronics, Peking University, Beijing 100871, China\\
\authormark{5}State Key Laboratory of Membrane Biology, Beijing Key Laboratory of Cardiometabolic Molecular Medicine, Institute of Molecular Medicine, Center for Life Sciences, College of Future Technology, Peking University, Beijing 100871, China\\
\authormark{6}School of Instrumentation and Optoelectronic Engineering, Beihang University, Beijing 100191, China\\
\authormark{7}Key Laboratory of RNA Biology, Institute of Biophysics, Chinese Academy of Sciences, Beijing, 100101, China\\
\authormark{8}National Laboratory of Biomacromolecules, Institute of Biophysics, Chinese Academy of Sciences, Beijing 100101, China\\
\authormark{9}PKU-IDG/McGovern Institute for Brain Research, Beijing 100871, China\\
\authormark{10}Beijing Academy of Artificial Intelligence, Beijing 100871, China\\
\authormark{11}Center for Quantitative Biology, Peking University, Beijing 100871, China\\
\authormark{12}Beijing Advanced Innovation Center for Imaging Theory and Technology, Capital Normal University, Beijing 100871, China\\
\authormark{13}heng.mao@pku.edu.cn\\
\authormark{$\dagger$}These authors contributed equally to this work\\
\authormark{*}Corresponding author:\email{ffeng@pku.edu.cn}}


\begin{abstract*} 
This article presents an approach for space-variant Shack-Hartmann wavefront reconstruction based on image moment theory. We derive the relation between the moment of a pair of subimages and the local transformation coefficients. A special solution is obtained from this relation using the square guide ‘star’. Our moment-based wavefront reconstruction has reduced computational complexity compared to the iteration-based algorithm. We execute image restorations using both the tiling strategy with 5 $\times$ 5 PSFs and the conventional strategy with a global average PSF. Our approach is supported by visual and quantitative evaluations.   
\end{abstract*}

\section{Introduction}
The problem of space-variant imaging is common in various microscopies, mostly due to random sample-induced aberrations. Several methods have been proposed to address this issue, including local point spread function (PSF) approximation \cite{Denis2015}, PSF interpolation \cite{Hirsch2010}, FISTA \cite{Beck2009}, and its variations \cite{Liang2022}. However, these methods all assume that the measurement matrix is known, and their customized optimizations for different experimental systems leave a significant gap with great potential.

In terms of the space-variant wavefront sensing problem, the most commonly used method is the Shack-Hartmann sensor (SH) with a point-like guide star. K. Wang et al. \cite{Wang2014} generated a nonlinear fluorescence guide star based on the two-photon effect from each isoplanatic zone. Then, the wavefronts from different positions or areas were reconstructed region by region \cite{Rajaeipour2020}. This method required multiple measurements and was stuck in the tradeoff between spatial resolution and acquisition speed. In addition, fluorescence beads spreading inside the sample produce their aberrated images, and the phase retrieval algorithm was used to restore the phases. A sophisticated algorithm is crucial since the fluorescence intensity of beads is very weak. In addition, D. Ancora et al \cite{Ancora2021} used a small circular mask on the pupil to separate superstitions of light from different directions and equivalently made a lenslet. Eighteen images are acquired each after rotations and scans to let the circle fulfill the pupil as much as possible. Inconveniences are obvious since multiple images are required and the pupil sampling resolution is also not satisfying.

In particular, recent works \cite{Wu2022,Wu2021} from the Qionghai Dai group have set a milestone by beautifully combining the digital adaptive optics technique (digital AO, DAO) with light-field microscopy. Due to the intrinsic advantages of self-provided microlenses in the system, space-variant wavefront sensing is naturally achieved from light-field images based on the correlation of each segmentation pair. The reconstructed wavefront then guides the light-field deconvolution, pushing light-field microscopy to the next stage. However, DAO is more suitable for applications with sufficient photons and high signal-to-noise-ratio images and is highly customized for light-field microscopy, limiting its benefits for other microscopes such as confocal, two-photon, light sheet, super-resolution, and bright field microscopes. In addition to pupil AO systems and their variants \cite{Park2017}, conjugate AO \cite{Mertz2015} is another popular method for addressing sample-induced aberration by targeting the plane near the sample where the space-variant wavefront originates. However, conjugate AO suffers from model approximation error, where sample-induced aberration is difficult to model with one phase screen and system aberration is not considered in this scheme.

We recently reported on a space-variant wavefront sensing scheme using the Shack-Hartmann sensor \cite{Feng2022}. The local space-variant function is characterized by six coefficients of the affine transformation, which are obtained through the optimization of the image similarity metric during registration. However, this algorithm requires multiple iterations and may not be suitable for microscopies requiring high-speed processing. In this study, we utilized a geometric transformation with five variables, excluding the shear transformation from the affine transformation and leaving only translation, scale, and rotation. The transformation coefficients were directly determined using the moment of images with an explicit formula, significantly reducing computing time. Our research has resulted in two notable findings that we shared in this paper. Our theory was validated using a widefield microscope, and our results for wavefront reconstruction and primitive image restoration support our approach. One significant benefit of our scheme is that only one aberrated SH image is required to resolve aberration across the field of view as far as the extended source, although a reference is currently needed.

\section{Theory}

The Shack-Hartmann sensor is widely used for wavefront sensing due to its high compatibility with various light sources, including point or extended sources and monochrome or colorful sources. It consists of a microlens array and a camera. A plane wavefront forms a standard subimage array on the camera, while an aberrated wavefront produces a disordered subimage array. The drift of the subimages indicates the gradient of the local wavefront \cite{Nightingale2013}. The entire wavefront can be reconstructed from the map of gradients using the assumption of continuity of the incident wavefront \cite{Soldevila2018} When a space-variant wavefront is present, the subimages in Shack-Hartmann sensors deform rather than merely shifting \cite{Feng2022}. In this paper, we propose using the moments of subimages to represent this deformation.

Image moments can represent specific characteristics, such as position, size, and direction \cite{Feng2018}. The origin and central moment of the image  $f\left( {x,y} \right)$ are defined as
\begin{equation}
    {O_{pq}} = \int {{x^p}} {y^q}f(x,y)dxdy,
    \label{eq:origin moment}
\end{equation}
\begin{equation}
    {C_{pq}} = \int {{{(x - {O_{10}})}^p}} {(y - {O_{01}})^q}f(x,y)dxdy.
    \label{eq:central moment}
\end{equation}
where $\left( x,y \right)$  is the lateral normalized coordinate of the image plane and the subscripts $pq$  indicate the order of moment in each direction. Our goal is to determine the transformation coefficients of these subimages directly from their moments. However, the general relationship between the coefficients and moments is a multivariate system of higher-order equations that does not support our objective. As a result, we are currently considering characteristic solutions for specific cases. In this manuscript, we use a square as the guide ‘star’ to significantly reduce the complexity of the coefficient-moment relationship. Due to the symmetry of squares, the odd central moments of squares are zero, and even moments of the same order are equal.

Furthermore, a geometrical transformation with five coefficients is utilized.
\begin{equation}
\left[ {\begin{array}{*{20}{c}}
{{x_2}}\\
{{y_2}}\\
1
\end{array}} \right] = \left[ {\begin{array}{*{20}{c}}
{{a_1}\cos \left( \theta  \right)}&{{a_1}\sin \left( \theta  \right)}&{{b_1}}\\
{ - {a_2}\sin \left( \theta  \right)}&{{a_2}\cos \left( \theta  \right)}&{{b_2}}\\
0&0&1
\end{array}} \right]\left[ {\begin{array}{*{20}{c}}
{{x_1}}\\
{{y_1}}\\
1
\end{array}} \right].
\end{equation}
These properties and simplifications significantly reduce the complexity of the equation system, resulting in a straightforward solution
\begin{equation}
\begin{small}
\begin{array}{l}
{a_1} = {\left( {\frac{{{J_{20}}}}{{{I_{20}}}}} \right)^{\frac{1}{2}}},\\
{a_2} = {\left( {\frac{{{J_{02}}}}{{{I_{02}}}}} \right)^{\frac{1}{2}}},\\
\theta  = \theta \left( {{J_{04}},{J_{13}},{J_{22}},{J_{31}},{J_{40}},{I_{40}},{I_{22}},{I_{04}}} \right),\theta  \in \left( { - \frac{\pi }{4},\frac{\pi }{4}} \right),\\
{b_1} = {J_{10}} - {a_1}\cos \left( \theta  \right){I_{10}} - {a_1}\sin \left( \theta  \right){I_{10}},\\
{b_2} = {J_{01}} + {a_2}\sin \left( \theta  \right){I_{01}} - {a_2}\cos \left( \theta  \right){I_{01}},
\end{array}
\end{small}
\label{eq:solution}
\end{equation}
where ${I_{10}},{I_{01}}$ and ${J_{10}},{J_{01}}$ represent the origin moments of subimages $I\left( {x,y} \right)$ and $J\left( {x,y} \right)$, ${I_{pq}},{J_{pq}},p + q > 1,p \in \mathbb{Z},q \in \mathbb{Z}$ represent  their central moments; $I\left( {x,y} \right)$ and $J\left( {x,y} \right)$ are the subimages without and with the aberration, respectively, or before and after the deformation. Once the local transformations of all lenslets have been determined, the space-variant wavefront can be reconstructed using the same scheme in Ref. \cite{Feng2022}. The complete derivations and simplifications from the multivariate system of higher-order equations to Eq.\eqref{eq:solution} and wavefront reconstruction are presented in the Appendix.

The explanation for Eq.\eqref{eq:solution} is intuitive. First, image stretching, which is characterized by second-order central moments, corresponds to the scaling factor of the geometric transformation. Image rotation, which is mainly determined by mixed moments, is expressed by the rotation factor. Finally, image shifting, which is led by first-order origin moments, is related to the translation factor.

\section{Experiments}
\subsection{Experimental setup}

\begin{figure*}[!htp]
\centering
\includegraphics[width=\linewidth]{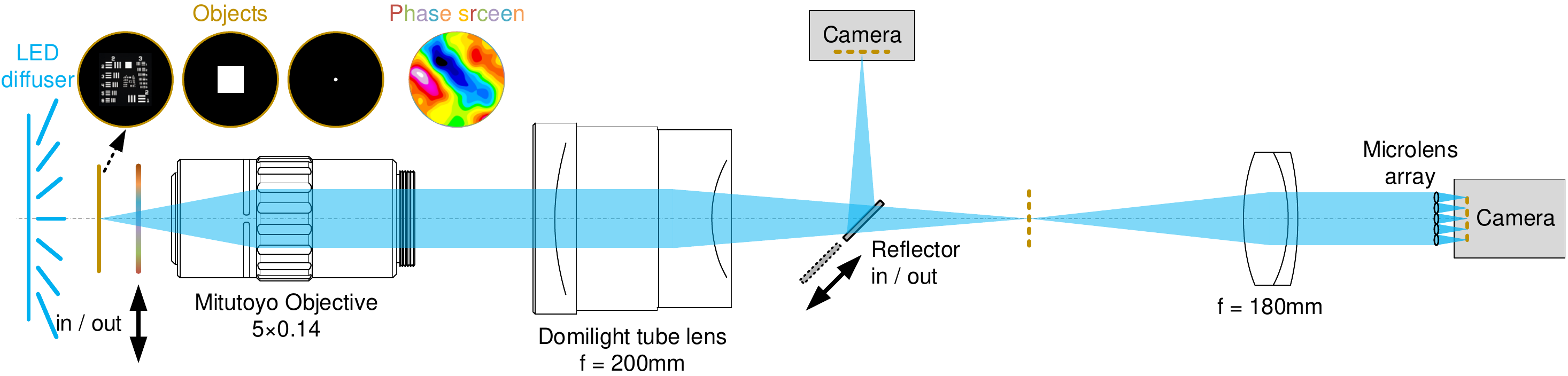}
\caption{Experimental setup}
\label{fig:Experimental setup}
\end{figure*}

As shown in Fig. \ref{fig:Experimental setup}, our experimental setup primarily consists of a microscope system with an objective (Mitutoyo, 5$\times$0.14, working distance 34 mm) and a tube lens (Domilight, focal length 200 mm). There are two arms in the detection space: one connects to a camera (FLIR, BFS-U3-16S2M-CS, pixel size 3.45 $\times$ 3.45 \textmu m$^2$, pixel number 1440 $\times$ 1080) for imaging, while the other connects to a lens (Thorlabs, AC508-180-A, focal length 180 mm) and a commercial Shack-Hartmann sensor (Optocraft, SHR4-130-GE, pitch 130 \textmu m, focal length 3.345 mm, pixel size 3.45 $\times$ 3.45 \textmu m$^2$, pixel number 3008 $\times$ 3000) for measuring the space-variant wavefront. The Shack-Hartmann sensor is conjugated to the back pupil plane of the objective. A reflector (Thorlabs, PF10-03-G01) installed on a flipper is used to switch the light between the imaging and sensing arms. The object, with an external diameter of 1 inch (24.5 mm), is located on the nominal plane of the objective. Three objects - a negative resolution target (Thorlabs, R1DS1N - 1951 USAF), a square mask (customized, side length 745 \textmu m), and a pinhole (customized, diameter 50 \textmu m) - are mounted on a runner and imaged and measured by the corresponding camera and Shack-Hartmann sensor without and with the phase screen, respectively. Additionally, the runner is mounted on a three-axis stage (x/y/z) for fine focusing since the objective is fixed. An LED diffuser panel is placed approximately 15 mm away from the object plane to create Köhler illumination.

In addition, the pinhole is translated to 3 $\times$ 3 positions with an interval of 372.6 \textmu m in each direction to measure the wavefront at corresponding locations. An acrylic plate, used as a phase screen to produce a space-variant wavefront, is placed between the object plane and the objective. The plate is heated by a hot air gun and presents a twisted surface. The square mask is used to perform our moment-based wavefront reconstruction, and its performance is evaluated by the reference wavefront using the conventional centroid-based method at 3 $\times$ 3 positions originating from pinhole illumination. The blurred image of the negative resolution target is used to implement space-variant image restoration. The central lenslet is intentionally blackened by its manufacturer Optocraft to block any light and leave a blank. The marginal pixels of raw 3008 $\times$ 3000 Shack-Hartmann images are cropped to 2901 $\times$ 2901 pixels, extremely close to the size of 77 $\times$ 77 pitches of lenslets, and then resized to 2849 $\times$ 2849 pixels using nearest-neighbor interpolation, where each lenslet is assigned 37 $\times$ 37 pixels. As a result, the back pupil of the objective is observed in the proximal inner circle of the square microlens array. Wavefront reconstructions are programmed in MATLAB® R2022b and run on a laptop with primary configurations of a 12th Intel i7-1260P CPU, 16 GB RAM and Windows 11 Professional operating system.

\subsection{Experimental results}

\begin{figure}[!htp]
\centering
\includegraphics[width=9cm]{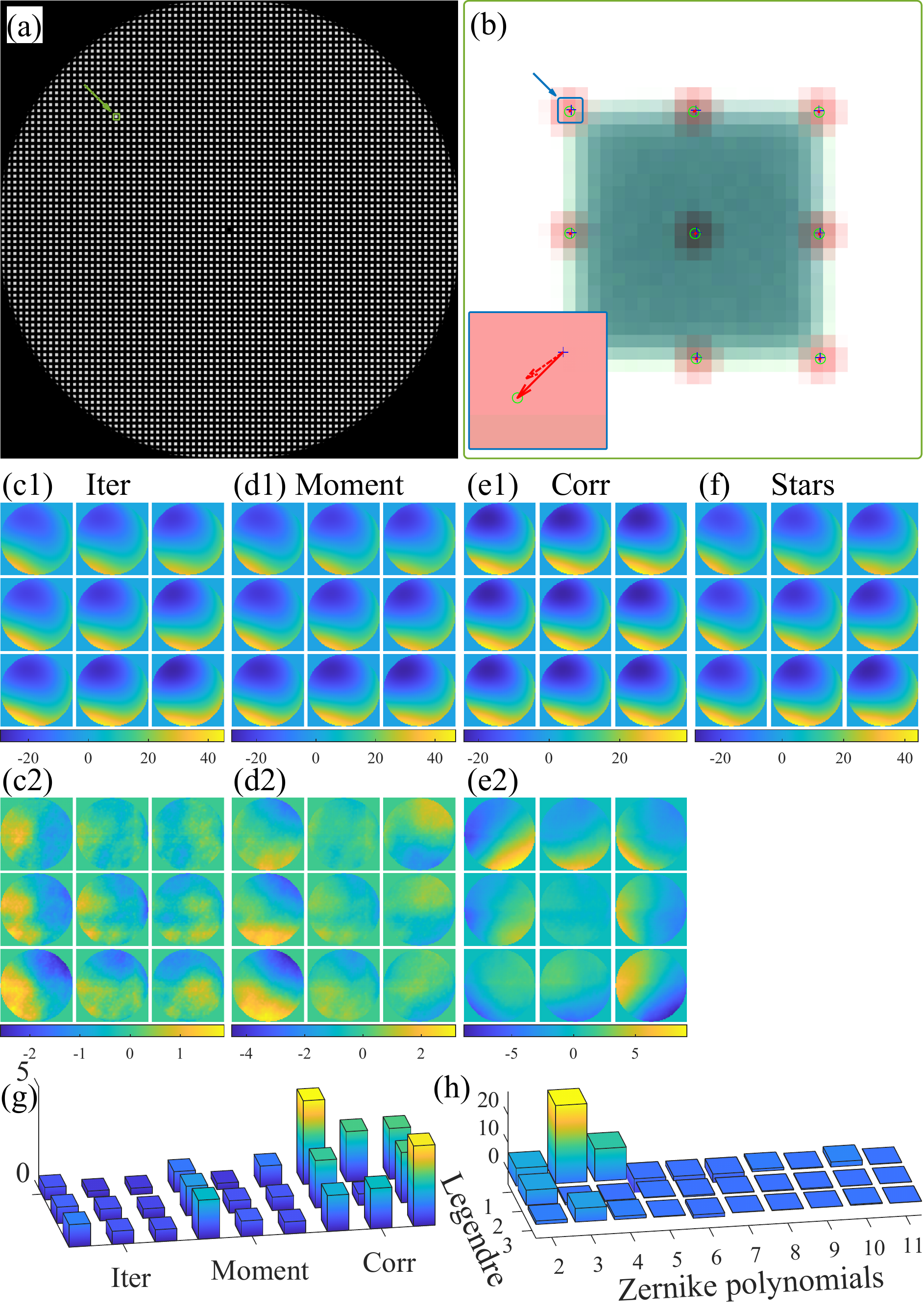} 
\caption{Experimental space-variant wavefront reconstructions, all scaled in radians. (a) Captured Shack-Hartmann image with phase screen. (b) Subimages of a square from the zoomed-in section in the box of (a), with an overlaid star array. Centroid shifts are indicated by a solid red arrow and compared with shifts predicted by transform estimation, indicated by a dotted red arrow. (c) Reconstruction wavefronts using (c1) the iteration algorithm, (d1) our moment-based method, (e1) the correlation method with a square source illuminated, and (f) the centroid-based method with 3 $\times$ 3 stars illuminated. (c2-e2) Residual wavefronts between (c1-d1) and (e). (g) RMS of the corresponding residual wavefronts (c2-d2). (h) Reconstructed Legendre-Zernike dual coefficients using our proposed moment-based method.}
\label{fig:Experimental results}
\end{figure}

Figure \ref{fig:Experimental results}(a) shows the SH image under square illumination with a phase screen. Two subimages of the square and one of the 3 $\times$ 3 star arrays with a phase screen are overlaid in Fig. \ref{fig:Experimental results}(b), showing 3 $\times$ 3 shifts calculated by the point centroid and the same predictions by our moment-based method from square deformation. The difference in shifts between the two methods is quite small. The Legendre-Zernike dual coefficients are reconstructed from the map of geometric transform coefficients, as shown in Fig. \ref{fig:Experimental results}(h). Four approaches—the iteration-based method, our moment-based method, the correlation method, and the centroid-based method as a reference—provide reconstructed wavefronts, as shown in Figures \ref{fig:Experimental results}(c1,d1,e1,f). The iterative algorithm optimizes the image similarity metric. The correlational method computes correlation coefficients from the normalized cross-correlation of the matrices. Note that the correlation method is space-invariant, so their nine wavefronts are identical.

The root mean square (RMS) of centroid-based wavefront reconstruction from the star array varies from 13.468 to 19.078 radians with an average of 16.082 radians within different views, and the peak-to-valley value (PV) is 55.388 to 72.250 radians with an average of 62.897 radians. As a result, residual wavefronts and their RMS are presented in Fig. \ref{fig:Experimental results}(c2,d2,e2) and Fig. \ref{fig:Experimental results}(g). The RMSs of wavefront reconstruction residuals using our moment-based method vary from 0.307 to 1.922 radians with an average of 0.927 radians within different views; those using the iteration-based registration algorithm vary from 0.306 to 1.104 radians with an average of 0.605 radians; and those using the correlation method vary from 0.562 to 4.116 radians with an average of 2.463 radians.

Methods based on the space-variant model, including moment-based and iteration-based algorithms, significantly outperform the correlation method within the space-invariant model. For the best estimation of one field of view (FOV), the performances of the moment-based and iteration-based algorithms are very close. When taking the average of all views, the root mean square (RMS) of the iteration-based algorithm is 1/3 smaller than that of the moment-based algorithm. However, the moment-based algorithm dramatically reduces computational complexity compared to the iteration-based algorithm. In our case, it takes only 0.942 seconds to calculate 4568 sets of coefficients in each lenslet in Fig. \ref{fig:Experimental results}(a) using the moment-based algorithm, while it takes 15.636 seconds to retrieve the same amount of data using the iteration-based algorithm. The advantage of computational complexity and time cost of the moment-based algorithm balances its disadvantage in accuracy of space-variant function reconstruction. This method would be useful for applications requiring high-speed processing, such as real-time closed-loop adaptive optics systems.

Upon closer observation, the residuals of the correlation method present opposite tip/tilt changes in each direction and lack a high-frequency component. These results imply that this space-variant wavefront mainly varies on slopes. The reconstructed Legendre-Zernike dual coefficients, shown in Fig. \ref{fig:Experimental results}(h), explain this phenomenon as expected. Zernike terms 2 and 3, which represent the tip/tilt of the wavefront in the pupil plane, change linearly since Legendre terms 2 and 3 are also ‘tip/tilt’ of the space-variant function in the image plane. However, while the result of the central view presents excellent accuracy of reconstruction using the correlation method, other views do not. In comparison, the results of every view using our method show consistency with the reference, indicating that the performance of our method is clearly superior to space-invariant reconstructions in space-variant applications.

\begin{figure}[!htp]
\centering
\includegraphics[width=9cm]{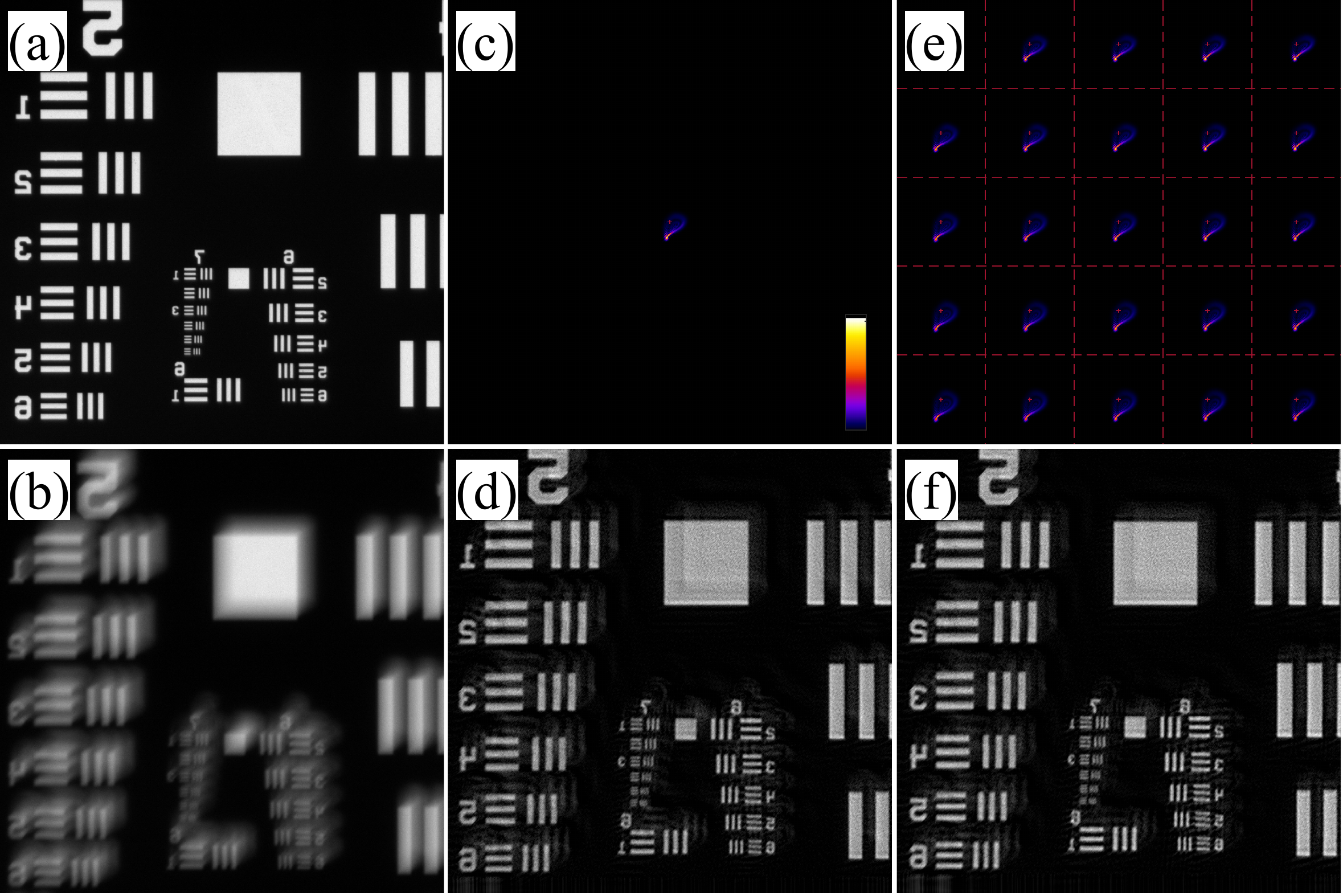}
\caption{Results of image restoration. (a-b) Images captured without and with the phase screen, respectively. (c) PSF corresponding to the reconstructed wavefront using the correlation method. (d) Results of image restoration using Wiener deconvolution based on (c) and (b). (e) PSFs corresponding to the reconstructed wavefronts using the moment-based method. (f) Results of image restoration after assembling each patch using Wiener deconvolution based on (e) and (b), tile by tile.}
\label{fig:Image restoration}
\end{figure}

To further address the space-variant problem, we provide a small demonstration of its importance and necessity in image restoration. The USAF target is imaged by the widefield camera without and with the same phase screen, as shown in Fig. \ref{fig:Image restoration}(a) and \ref{fig:Image restoration}(b). The wavefront we reconstructed can be used to restore details and features from the blurred image using an algorithm known as digital adaptive optics. For space-invariant restoration, the wavefront obtained using the correlation method, shown in Fig. 2(e1), is used to generate the global average PSF, as shown in Fig. \ref{fig:Image restoration}(c). The blurred image captured with the phase screen, shown in Fig. \ref{fig:Image restoration}(b), is deblurred using the Wiener filter with this PSF, resulting in Fig. \ref{fig:Image restoration}(d). For space-variant restoration, the wavefronts obtained using our moment-based method are used to generate PSFs at each center of 5 $\times$ 5 tiles, as shown in Fig. \ref{fig:Image restoration}(e). The blurred image is deblurred using Wiener filters with all 5 $\times$ 5 PSFs. The final result, shown in Fig. \ref{fig:Image restoration}(f), is assembled from the corresponding tiles. During Wiener deconvolution, the raw 1080 $\times$ 1080 image with a centered region of interest (ROI) is first extended to 1479 $\times$ 1479 using a smooth extension method of order 0 and then cropped to 1080 $\times$ 1080 with a consistent center after deconvolution to improve image edge quality. Visual observations clearly show better restoration using our scheme in terms of position, contrast, resolution, and artifacts.

Quantitative evaluations using frequently used indexes such as structural similarity (SSIM) \cite{Zhou2004}, correlation (R) \cite{Fisher1992}, peak signal-to-noise ratio (PSNR) \cite{Huynh2012}, multiscale structural similarity (MULTISSIM) \cite{Wang2003}, and mean-squared error (MSE) \cite{Gonzalez2018} are made between Fig. \ref{fig:Image restoration}(b,d,f) and Fig. \ref{fig:Image restoration}(a), as listed in Table \ref{tab:Evaluations}. The higher the first four indexes are, the closer the image is to the unperturbed image. Additionally, the lower the MSE is, the better. As seen from the results, our proposal outperforms the conventional correlation method. It is worth mentioning that the SSIM of the blurred image in Fig. \ref{fig:Image restoration}(b) is higher than that of the restored images in Fig. \ref{fig:Image restoration}(c,d), which is counterintuitive. A possible explanation is that every evaluation index has limitations, and this case happens to reach this limit \cite{Nilsson2020}.

\begin{table}[ht]
\centering
\caption{Image quality evaluations between Fig. \ref{fig:Image restoration}(b, d, f), as blurred, corr, moment, and Fig. \ref{fig:Image restoration}(a) the ground truth.}
\begin{tabular}{ccccccc}
\hline
 & SSIM & R & PSNR & MultiSSIM & MSE \\
\hline
  Blurred & 0.549 & 0.698 & 13.396 & 0.569 & 0.046 \\ 
  Corr & 0.407 & 0.868 & 16.553 & 0.699 & 0.022 \\ 
  Moment & 0.479 & 0.959 & 21.289 & 0.809 & 0.007 \\ 
\hline
\end{tabular}
 \label{tab:Evaluations}
\end{table}

In this study, image restoration was implemented using the MATLAB platform. The PSF was calculated using the fast Fourier transform (FFT) algorithm based on the Fourier transform property of a lens. -in functions such as ‘imregtform’, ‘deconvwnr’, ‘wextend’, ‘ssim’, ‘corr2’, ‘psnr’, ‘multissim’, and ‘immse’ were used for registration, deconvolution, extension, and evaluation.

It should be noted that the tiling method employed in the experiments served only to demonstrate the superiority and significance of the space-variant model for image restoration. Once the space-variant wavefront was obtained through our approaches, the measurement matrix $\bf{A}$ was determined, consisting of PSFs for each object point. Each PSF corresponded to its respective wavefront. The space-variant imaging forward model was thus presented as $\bf{b} = \bf{Ax}$, where $\bf{x}$ is the vector representing the true image and $\bf{b}$ is the vector representing the blurred image captured by the camera. In future work, it would be interesting to effectively combine restoration algorithms such as PSF approximation \cite{Denis2015} and FISTA \cite{Beck2009} with our space-variant wavefront reconstruction.

\section{Discussion}

\begin{figure}[htbp!]
\centering
\includegraphics[width=9cm]{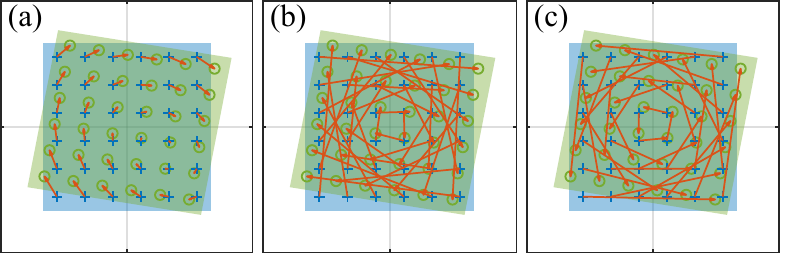}
\caption{Rotation ambiguity with the same pair of squares. (a) Measurable rotation. (b) Rotation angle of (a) plus 90°. (c) Rotation angle of (a) minus 90°.}
\label{fig:Rotation ambiguity}
\end{figure}

Using a square as the illumination source results in an obvious drawback of rotation ambiguity. Specifically, the system of equations for moments and geometric coefficients has multiple solutions for the rotation angle when squares are used, as described in Eq. \eqref{eq:resultion rotation with ambiguity}. Figure \ref{fig:Rotation ambiguity} presents a schematic example with three physically possible and meaningful displacement field situations, among many others. Intuitively, solutions with smaller absolute values of the rotation angle are preferred. Additionally, displacements characterize the power required to move a point from one location to another. Different displacement fields imply different total powers needed to achieve the same image deformation. The principle of minimum power is implemented and preferred among all solutions. In practice, the solution with the smallest absolute value of the rotation angle is chosen, as shown in Fig. \ref{fig:Rotation ambiguity}(a). As a result, the measurable range of the rotation angle for our moment-based method is limited to (-45°,45°). The aberration leading to subimage rotation is spatially spiral, resembling the vortex beam. The difference lies in whether the domain is spiral - either the space-variant function or the pupil function.

In principle, equilateral geometric figures such as triangles, pentagons, hexagons, heptagons, octagons, etc., or point arrays could be used as guide ‘stars’ with different forms of solutions. However, there are three advantages to using squares. First, square subimages naturally fill the FOV of the lenslet on the camera, providing uniform contributions to space-variant wavefront sensing. Other figures with unfilled regions could also be used for sensing, but their performance is inferior to that of squares. Second, the subarea of a square lenslet with equivalent unfilled marginal space provides an isotropic dynamic range of square translation for wavefront sensing in different directions. Finally, the dynamic range of the square’s rotation angle is large, up to 90°. In comparison, the dynamic ranges for equilateral triangles, pentagons, hexagons, octagons, etc., are 120°, 72°, 60°, 51.42° and 45°, respectively. Among these figures, squares outperform most of them except for triangles.

The results indicate that the RMSs of residual wavefronts obtained using the moment-based algorithm are inferior to those obtained using the iteration-based algorithm. There are two main reasons for this outcome. First, the moment-based algorithm adopts a five-coefficient geometrical transformation for resolution, while the iteration-based algorithm uses a six-coefficient affine transformation. The degree of freedom in the moment-based algorithm is inherently deficient, and a higher degree would better characterize the delicate features of the space-variant function. Second, the moment-based algorithm assumes that the image of a standard square is deformed. However, in practical systems, SH subimages without phase screens may not be standard due to assembly errors, manufacturing defects, and image noise. It is important to note that the space-variant function is an arbitrary function and should be modeled using a combination of polynomials. However, this paper adopts a linear model where second-order and higher items are neglected. The presence of higher items in experiments produces model error, resulting in compromised accuracy of aberration measurement. The robustness of our method primarily depends on this model error. In other words, the space-variant function is not severe as long as the FOV is not too large or the tissue depth is not too deep.

\section{Conclusion}
In conclusion, we have proposed a moment-based algorithm for estimating transformation coefficients in the space-variant Shack-Hartmann wavefront reconstruction scheme. Two main contributions are provided. First, the relation between the transformation coefficients and image moments is derived, and a special solution with squares is presented for the first time. Second, a special solution is used to directly reconstruct the space-variant wavefront and significantly improve the calculation speed. A microscope system was implemented to demonstrate the effectiveness of our approach. Experimental results for wavefront reconstruction and image restoration using our moment-based method showed a nearly 15-fold improvement in processing speed and a 30$\%$ compromise in the RMS of residual wavefronts on average compared to the iteration-based algorithm.

\section*{Appendix}

\subsection*{Special solution for transformation coefficients from image moments}
The image captured by the camera usually has been normalized
\begin{equation}
    f\left( {x,y} \right) = \frac{{\mathord{\buildrel{\lower3pt\hbox{$\scriptscriptstyle\frown$}} 
\over f} \left( {x,y} \right)}}{{\int {\mathord{\buildrel{\lower3pt\hbox{$\scriptscriptstyle\frown$}} 
\over f} \left( {x,y} \right)dxdy} }},
\end{equation}
where $\left( {x,y} \right)$ is the lateral normalized coordinate of the image plane, $\mathord{\buildrel{\lower3pt\hbox{$\scriptscriptstyle\frown$}} 
\over f} \left( {x,y} \right)$ are the raw images, and $f\left( {x,y} \right)$ are the corresponding normalized images. The normalization makes the image a probability density distribution since the pixel value of the image is positive. The definitions of the origin and central moment are rewritten here:
\begin{equation}
    {O_{pq}} = \int {{x^p}} {y^q}f(x,y)dxdy,
    \label{eq:origin moment 2}
\end{equation}
\begin{equation}
    {C_{pq}} = \int {{{(x - {O_{10}})}^p}} {(y - {O_{01}})^q}f(x,y)dxdy.
    \label{eq:central moment 2}
\end{equation}

The affine transformation with six coefficients has been used to characterize the deformation of a pair of subimages
\begin{equation}
 \left[ {\begin{array}{*{20}{c}}
{{x_2}}\\
{{y_2}}\\
1
\end{array}} \right] = \left[ {\begin{array}{*{20}{c}}
{{a_{11}}}&{{a_{12}}}&{{b_1}}\\
{{a_{21}}}&{{a_{22}}}&{{b_2}}\\
0&0&1
\end{array}} \right]\left[ {\begin{array}{*{20}{c}}
{{x_1}}\\
{{y_1}}\\
1
\end{array}} \right].
\label{eq:affine transformation}
\end{equation}
Let $I\left( {x,y} \right)$ and $J\left( {x,y} \right)$ be a pair of images without and with aberration or before and after transformation. ${I_{10}},{I_{01}}$ and ${J_{10}},{J_{01}}$ denote the corresponding origin moment, ${I_{pq}},{J_{pq}},p + q > 1,p \in \mathbb{Z},q \in \mathbb{Z}$ denote corresponding central moment, and the subscripts $pq$ indicate the order of moment in each direction. From Eq.\eqref{eq:affine transformation}, the relation between the coordinate systems of the images is
\begin{equation}
\begin{array}{l}
{x_2} = {a_{11}}{x_1} + {a_{12}}{y_1} + {b_1},\\
{y_2} = {a_{21}}{x_1} + {a_{22}}{y_1} + {b_2}.
\end{array}
\label{eq:coordinate systems}
\end{equation}
Substituting Eq.\eqref{eq:coordinate systems} to the formula of Eq.\eqref{eq:origin moment 2} gives \cite{Pei1995}
\begin{equation}
\begin{aligned}
{J_{10}} &= \int_\Sigma  {{x_2}J({x_2},{y_2})d{x_2}d{y_2}} \\
 &= \int_\Omega  {\left( {{a_{11}}{x_1} + {a_{12}}{y_1} + {b_1}} \right)I({x_1},{y_1})d{x_1}d{y_1}} \\
 &= {a_{11}}{I_{10}} + {a_{12}}{I_{01}} + {b_1}
\end{aligned}.
\end{equation}
Following similar deductions, we will have the raw relation between the image moment and the affine transformation coefficients,
\begin{equation}
\begin{array}{l}
{J_{10}} = {a_{11}}{I_{10}} + {a_{12}}{I_{01}} + {b_1},\\
{J_{01}} = {a_{21}}{I_{10}} + {a_{22}}{I_{01}} + {b_2},\\
\\
{J_{20}} = {a_{11}}^2{I_{20}} + {a_{12}}^2{I_{02}} + 2{a_{11}}{a_{12}}{I_{11}},\\
{J_{11}} = {a_{11}}{a_{21}}{I_{20}} + {a_{12}}{a_{22}}{I_{02}} + ({a_{11}}{a_{22}} + {a_{12}}{a_{21}}){I_{11}},\\
{J_{02}} = {a_{21}}^2{I_{20}} + {a_{22}}^2{I_{02}} + 2{a_{21}}{a_{22}}{I_{11}},\\
\\
{J_{30}} = {a_{11}}^3{I_{30}} + 3{a_{11}}^2{a_{12}}{I_{21}} + 3{a_{11}}{a_{12}}^2{I_{12}} + {a_{12}}^3{I_{03}},\\
{J_{21}} = {a_{11}}^2{a_{21}}{I_{30}} + ({a_{11}}^2{a_{22}} + 2{a_{11}}{a_{12}}{a_{21}}){I_{21}} + ({a_{21}}{a_{12}}^2 + 2{a_{12}}{a_{11}}{a_{22}}){I_{12}} + {a_{12}}^2{a_{22}}{I_{03}},\\
{J_{12}} = {a_{11}}{a_{21}}^2{I_{30}} + ({a_{12}}{a_{21}}^2 + 2{a_{11}}{a_{22}}{a_{21}}){I_{21}} + ({a_{11}}{a_{22}}^2 + 2{a_{12}}{a_{21}}{a_{22}}){I_{12}} + {a_{12}}{a_{22}}^2{I_{03}},\\
{J_{03}} = {a_{21}}^3{I_{30}} + 3{a_{21}}^2{a_{22}}{I_{21}} + 3{a_{21}}{a_{22}}^2{I_{12}} + {a_{22}}^3{I_{03}},\\
\\
{J_{40}} = {a_{11}}^4{I_{40}} + 4{a_{11}}^3{a_{12}}{I_{31}} + 6{a_{11}}^2{a_{12}}^2{I_{22}} + 4{a_{11}}{a_{12}}^3{I_{13}} + {a_{12}}^4{I_{04}},\\
{J_{31}} = {a_{11}}^3{a_{21}}{I_{40}} + ({a_{11}}^3{a_{22}} + 3{a_{11}}^2{a_{12}}{a_{21}}){I_{31}} + (3{a_{11}}^2{a_{12}}{a_{22}} + 3{a_{11}}{a_{12}}^2{a_{21}}){I_{22}} + \\
\qquad\qquad\qquad\qquad\qquad\qquad\qquad\qquad\qquad\qquad({a_{21}}{a_{12}}^3 + 3{a_{11}}{a_{12}}^2{a_{22}}){I_{13}} + {a_{12}}^3{a_{22}}{I_{04}},\\
{J_{22}} = {a_{11}}^2{a_{21}}^2{I_{40}} + (2{a_{11}}^2{a_{22}}{a_{21}} + 2{a_{11}}{a_{12}}{a_{21}}^2){I_{31}} + ({a_{11}}^2{a_{22}}^2 + 4{a_{11}}{a_{12}}{a_{21}}{a_{22}} + {a_{12}}^2{a_{21}}^2){I_{22}}  \\
\qquad\qquad\qquad\qquad\qquad\qquad\qquad\qquad
+(2{a_{21}}{a_{12}}^2{a_{22}} + 2{a_{11}}{a_{12}}{a_{22}}^2){I_{13}} + {a_{12}}^2{a_{22}}^2{I_{04}},\\
{J_{13}} = {a_{11}}{a_{21}}^3{I_{40}} + ({a_{12}}{a_{21}}^3 + 3{a_{11}}{a_{22}}{a_{21}}^2){I_{31}} + (3{a_{11}}{a_{21}}{a_{22}}^2 + 3{a_{12}}{a_{21}}^2{a_{22}}){I_{22}} + \\
\qquad\qquad\qquad\qquad\qquad\qquad\qquad\qquad\qquad\qquad({a_{11}}{a_{22}}^3 + 3{a_{12}}{a_{21}}{a_{22}}^2){I_{13}} + {a_{12}}{a_{22}}^3{I_{04}},\\
{J_{04}} = {a_{21}}^4{I_{40}} + 4{a_{21}}^3{a_{22}}{I_{31}} + 6{a_{21}}^2{a_{22}}^2{I_{22}} + 4{a_{21}}{a_{22}}^3{I_{13}} + {a_{22}}^4{I_{04}}.
\end{array}
\label{eq:general system}
\end{equation}
This multivariate system of higher-order equations has difficulty resolving affine coefficients $\left\{ {{a_{11}},{a_{21}},{a_{21}},{a_{22}},{b_1},{b_2}} \right\}$ from $\left\{ {{I_{pq}},{J_{pq}}} \right\}$. Proper simplifications are required to address this problem. In this paper, the five-coefficient geometrical transformation is used instead of the six-coefficient affine transformation, which is presented as
\begin{equation}
\left[ {\begin{array}{*{20}{c}}
{{x_2}}\\
{{y_2}}\\
1
\end{array}} \right] = \left[ {\begin{array}{*{20}{c}}
{{a_1}\cos \left( \theta  \right)}&{{a_1}\sin \left( \theta  \right)}&{{b_1}}\\
{ - {a_2}\sin \left( \theta  \right)}&{{a_2}\cos \left( \theta  \right)}&{{b_2}}\\
0&0&1
\end{array}} \right]\left[ {\begin{array}{*{20}{c}}
{{x_1}}\\
{{y_1}}\\
1
\end{array}} \right].
\label{eq:geometrical transformation}
\end{equation}
At the same time, the square image is used for the guide ‘star’. Its moments have so many properties
\begin{equation}
{I_{11}} = 0, {I_{31}} = 0, {I_{13}} = 0, {I_{20}} = {I_{02}}, {I_{40}} = {I_{04}}.
\label{eq:moments simplifications}
\end{equation}
Substituting Eqs.\eqref{eq:geometrical transformation} and \eqref{eq:moments simplifications} to Eq.\eqref{eq:general system} gives
\begin{equation}
\begin{array}{c}
{J_{20}} = a_1^2{I_{20}},\\
{J_{02}} = a_2^2{I_{02}}.
\end{array}
\end{equation}
Thus, the scaling factor is resolved
\begin{equation}
{a_1} = {\left( {\frac{{{J_{20}}}}{{{I_{20}}}}} \right)^{\frac{1}{2}}},{a_2} = {\left( {\frac{{{J_{02}}}}{{{I_{02}}}}} \right)^{\frac{1}{2}}}.
\label{eq:resultion scaling}
\end{equation}
In addition, from the fourth-order moments, we have
\begin{equation}
\begin{aligned}
\cos (4\theta ) = \frac{{{{4{J_{04}}} \mathord{\left/
 {\vphantom {{4{J_{04}}} {a_1^4}}} \right.
 \kern-\nulldelimiterspace} {a_1^4}} - 3{I_{22}} - 3{I_{40}}}}{{{I_{40}} - 3{I_{22}}}},\\
\cos (4\theta ) = \frac{{{{4{J_{22}}} \mathord{\left/
 {\vphantom {{4{J_{22}}} {a_1^2a_2^2}}} \right.
 \kern-\nulldelimiterspace} {a_1^2a_2^2}} - {I_{22}} - {I_{40}}}}{{3{I_{22}} - {I_{40}}}},\\
\cos (4\theta ) = \frac{{{{4{J_{40}}} \mathord{\left/
 {\vphantom {{4{J_{40}}} {a_2^4}}} \right.
 \kern-\nulldelimiterspace} {a_2^4}} - 3{I_{22}} - 3{I_{40}}}}{{{I_{40}} - 3{I_{22}}}},
\label{eq:cos4theta}
\end{aligned}
\end{equation}
\begin{equation}
\begin{aligned}
\sin (4\theta ) = \frac{{4{J_{13}}}}{{a_1^3{a_2}\left( {3{I_{22}} - {I_{40}}} \right)}},\\
\sin (4\theta ) = \frac{{4{J_{31}}}}{{{a_1}a_2^3\left( {{I_{40}} - 3{I_{22}}} \right)}}.
\label{eq:sin4theta}
\end{aligned}
\end{equation}
According to Euler's formula, we have
\begin{small}
\begin{equation}
\exp (\rm{i} 4\theta ) = \frac{{{{4{J_{04}}} \mathord{\left/
 {\vphantom {{4{J_{04}}} {a_1^4}}} \right.
 \kern-\nulldelimiterspace} {a_1^4}} + {{4{J_{40}}} \mathord{\left/
 {\vphantom {{4{J_{40}}} {a_2^4}}} \right.
 \kern-\nulldelimiterspace} {a_2^4}} - {{4{J_{22}}} \mathord{\left/
 {\vphantom {{4{J_{22}}} {a_1^2a_2^2}}} \right.
 \kern-\nulldelimiterspace} {a_1^2a_2^2}} - 5{I_{22}} - {{5\left( {{I_{40}} + {I_{04}}} \right)} \mathord{\left/
 {\vphantom {{5\left( {{I_{40}} + {I_{04}}} \right)} 2}} \right.
 \kern-\nulldelimiterspace} 2}}}{{3{{\left( {{I_{40}} + {I_{04}}} \right)} \mathord{\left/
 {\vphantom {{\left( {{I_{40}} + {I_{04}}} \right)} 2}} \right.
 \kern-\nulldelimiterspace} 2} - 9{I_{22}}}} + {\rm{i}}\frac{{{{4{J_{31}}} \mathord{\left/
 {\vphantom {{4{J_{31}}} {{a_1}a_2^3}}} \right.
 \kern-\nulldelimiterspace} {{a_1}a_2^3}} - {{4{J_{13}}} \mathord{\left/
 {\vphantom {{4{J_{13}}} {a_1^3{a_2}}}} \right.
 \kern-\nulldelimiterspace} {a_1^3{a_2}}}}}{{\left( {{I_{40}} + {I_{04}}} \right) - 6{I_{22}}}}.
 \label{eq:Euleformular}
\end{equation}
 \end{small}
The averages of $\cos (4\theta )$ and $\sin (4\theta )$ of Eqs.\eqref{eq:cos4theta} and \eqref{eq:sin4theta} are used in Eq.\eqref{eq:Euleformular}. Thus, the rotation factor is calculated as follows:
\begin{small}
\begin{equation}
\theta  = \frac{1}{4}{\rm{angle}}\left[ \frac{{{{4{J_{04}}} \mathord{\left/
 {\vphantom {{4{J_{04}}} {a_1^4}}} \right.
 \kern-\nulldelimiterspace} {a_1^4}} + {{4{J_{40}}} \mathord{\left/
 {\vphantom {{4{J_{40}}} {a_2^4}}} \right.
 \kern-\nulldelimiterspace} {a_2^4}} - {{4{J_{22}}} \mathord{\left/
 {\vphantom {{4{J_{22}}} {a_1^2a_2^2}}} \right.
 \kern-\nulldelimiterspace} {a_1^2a_2^2}} - 5{I_{22}} - {{5\left( {{I_{40}} + {I_{04}}} \right)} \mathord{\left/
 {\vphantom {{5\left( {{I_{40}} + {I_{04}}} \right)} 2}} \right.
 \kern-\nulldelimiterspace} 2}}}{{3{{\left( {{I_{40}} + {I_{04}}} \right)} \mathord{\left/
 {\vphantom {{\left( {{I_{40}} + {I_{04}}} \right)} 2}} \right.
 \kern-\nulldelimiterspace} 2} - 9{I_{22}}}} + {\rm{i}}\frac{{{{4{J_{31}}} \mathord{\left/
 {\vphantom {{4{J_{31}}} {{a_1}a_2^3}}} \right.
 \kern-\nulldelimiterspace} {{a_1}a_2^3}} - {{4{J_{13}}} \mathord{\left/
 {\vphantom {{4{J_{13}}} {a_1^3{a_2}}}} \right.
 \kern-\nulldelimiterspace} {a_1^3{a_2}}}}}{{\left( {{I_{40}} + {I_{04}}} \right) - 6{I_{22}}}} \right] + \frac{\pi }{2}k,k \in \mathbb{Z},
  \label{eq:resultion rotation with ambiguity}
\end{equation}
\end{small}
where angle means taking the phase angle from a complex number. For the rotation ambiguity, the principle of minimization energy is applied, and the smallest absolute value is taken
\begin{small}
\begin{equation}
\theta  = \frac{1}{4}{\rm{angle}}\left[ \frac{{{{4{J_{04}}} \mathord{\left/
 {\vphantom {{4{J_{04}}} {a_1^4}}} \right.
 \kern-\nulldelimiterspace} {a_1^4}} + {{4{J_{40}}} \mathord{\left/
 {\vphantom {{4{J_{40}}} {a_2^4}}} \right.
 \kern-\nulldelimiterspace} {a_2^4}} - {{4{J_{22}}} \mathord{\left/
 {\vphantom {{4{J_{22}}} {a_1^2a_2^2}}} \right.
 \kern-\nulldelimiterspace} {a_1^2a_2^2}} - 5{I_{22}} - {{5\left( {{I_{40}} + {I_{04}}} \right)} \mathord{\left/
 {\vphantom {{5\left( {{I_{40}} + {I_{04}}} \right)} 2}} \right.
 \kern-\nulldelimiterspace} 2}}}{{3{{\left( {{I_{40}} + {I_{04}}} \right)} \mathord{\left/
 {\vphantom {{\left( {{I_{40}} + {I_{04}}} \right)} 2}} \right.
 \kern-\nulldelimiterspace} 2} - 9{I_{22}}}} + {\rm{i}}\frac{{{{4{J_{31}}} \mathord{\left/
 {\vphantom {{4{J_{31}}} {{a_1}a_2^3}}} \right.
 \kern-\nulldelimiterspace} {{a_1}a_2^3}} - {{4{J_{13}}} \mathord{\left/
 {\vphantom {{4{J_{13}}} {a_1^3{a_2}}}} \right.
 \kern-\nulldelimiterspace} {a_1^3{a_2}}}}}{{\left( {{I_{40}} + {I_{04}}} \right) - 6{I_{22}}}} \right],
 \label{eq:resultion rotation}
\end{equation}
\end{small}
where $\theta  \in \left( { - \frac{\pi }{4},\frac{\pi }{4}} \right)$. A detailed discussion about rotation ambiguity is presented in the paper. The formula of first-order moments is rewritten as
\begin{equation}
\begin{array}{c}
{J_{10}} = {a_1}\cos \left( \theta  \right){I_{10}} + {a_1}\sin \left( \theta  \right){I_{10}} + {b_1},\\
{J_{01}} =  - {a_2}\sin \left( \theta  \right){I_{01}} + {a_2}\cos \left( \theta  \right){I_{01}} + {b_2}.
\end{array}
\label{eq:first-order moments}
\end{equation}
Since the scaling factor and rotation factor are both resolved, the shifting factor is presented directly from Eq.\eqref{eq:first-order moments}
\begin{equation}
\begin{array}{c}
{b_1} = {J_{10}} - {a_1}\cos \left( \theta  \right){I_{10}} - {a_1}\sin \left( \theta  \right){I_{10}},\\
{b_2} = {J_{01}} + {a_2}\sin \left( \theta  \right){I_{01}} - {a_2}\cos \left( \theta  \right){I_{01}}.
\end{array}
\label{eq:resultion shifting}
\end{equation}

In summary, a special solution of squares for transformation coefficients from image moments is obtained as Eqs.(\ref{eq:resultion scaling}, \ref{eq:resultion rotation}, \ref{eq:resultion shifting}). Following the same procedure in Ref \cite{Feng2022}, the dual-orthogonal coefficients could be reconstructed once the transformation coefficients are obtained.

\begin{backmatter}
\bmsection{Funding} National Natural Science Foundation of China (32071458, 21927813, 81827809, 81925022, 92054301, 92150301), National Key Research and Development Program of China (2022YFC3400600), Beijing Natural Science Foundation Key Research Topics (Z20J00059), Frontier Laboratory of Super-resolution Imaging in Jingjinji National Center of Technology Innovation.

\bmsection{Disclosures} The authors declare no conflicts of interest.
\end{backmatter}

\bibliography{sample}

\end{document}